\documentclass[amsmath,amssymb,twocolumn]{revtex4}

\usepackage{bm}
\usepackage{color}
\usepackage{dcolumn}
\usepackage{epsfig}
\usepackage{graphicx,rotating}
\usepackage{longtable}
\usepackage{mhchem}
\usepackage{gensymb}
\usepackage{multirow}
\usepackage{balance}
\usepackage[caption=false]{subfig}

\newcolumntype{C}[1]{>{\centering\arraybackslash}m{#1}} 

\begin{document}

\title{Ultralow and Anisotropic Thermal Conductivity in Semiconductor As$_2$Se$_3$}

\author{Robert L. Gonz\'alez-Romero}
\affiliation{Department of Physical, Chemical and Natural Systems, University Pablo de Olavide. Ctra. de Utrera, km. 1, 41013, Sevilla, Spain}
\author{Alex Antonelli, Anderson S. Chaves}
\affiliation{Instituto de F\'isica Gleb Wataghin and Centre for Computational Engineering \& Sciences, Universidade Estadual do Campinas, Campinas, Brazil}
\author{Juan J. Mel\'endez}
\affiliation{Department of Physics, University of Extremadura. Avda. de Elvas, s/n, 06006, Badajoz, Spain}
\affiliation{Institute for Advanced Scientific Computing of Extremadura, Badajoz, Spain}

\date{\today}

\begin{abstract}
An ultralow lattice thermal conductivity of 0.14 W$\cdot$ m$^{-1} \cdot$ K$^{-1}$ along the $\vec b$ axis of As$_2$Se$_3$ single crystals was obtained at 300 K by first-principles calculations involving the density functional theory and the resolution of the Boltzmann transport equation. This ultralow lattice thermal conductivity arises from the combination of two mechanisms: 1) a cascade-like fall of the low-lying optical modes, which results in avoided crossings of these with the acoustic modes, low sound velocities and increased scattering rates of the acoustic phonons; and 2) the repulsion between the lone-pair electrons of the As cations and the valence $p$ orbitals of the Se anions, which leads to an increase in the anharmonicity of the bonds. The physical origins of these mechanisms lie on the nature of the chemical bonding in the material and its strong anisotropy. These results, whose validity has been addressed by comparison with SnSe, for which excellent agreement between the theoretical predictions and the experiments is achieved, point out that As$_2$Se$_3$ could exhibit improved thermoelectric properties.
\end{abstract}

\maketitle

Semiconductor materials have played a crucial role in the development of the modern societies due to their conducting properties intimately related to the temperature. Lately, insofar the electronic devices are becoming smaller, besides their electronic and optical behavior, their thermal and thermoelectric performances are becoming more and more relevant. In particular, the thermoelectric materials have received considerable attention during the last decades because of their capability to convert waste-heat into electricity, thus improving the efficiency of many industrial processes. In this sense, semiconductors with low thermal conductivities are indispensable to design and develop high-efficiency thermoelectric devices \cite{Yang-16,He-15,Zhao-14}. Indeed, the thermoelectric efficiency of a material is given by its figure of merit $zT$, defined at a temperature $T$ as:
\begin{equation}
	zT = \frac {S^2\sigma}{\kappa}T,
	\label{eq:zt}
\end{equation}
where $S$ is the Seebeck coefficient and $\sigma$ and $\kappa$ are the electrical and thermal conductivity, respectively. The latter may be written in terms of the electronic $\kappa_e$ and lattice $\kappa_L$ contributions as:
\begin{equation}
	\kappa = \kappa_e + \kappa_L.
	\label{eq:kappa}
\end{equation}

The correlation between $\sigma$, $S$ and $\kappa$ has been largely discussed in the literature (see, for instance, Ref. \cite{Snyder-08}). For the purpose of this work, we will just mention that a low thermal conductivity is a desirable condition for thermoelectric performance. In addition, we also remark that, as a general rule, the lattice vibrations (phonons) drive most of the heat at room temperature or higher, and this transport is ruled mostly by the phonon-phonon scattering processes resulting from the anharmonicity of the interatomic potential. At low temperatures (or in nanostructured samples), extrinsic scattering by point defects, isotopes or grain boundaries become relevant as well \cite{Ziman-60}. For this reason, we will limit ourselves to study the lattice contribution to the thermal conductivity, $\kappa_L$. 

From the previous paragraphs, it seems clear that an accurate description and comprehension of the several phonon scattering processes is required to understand how material properties relate to $\kappa_L$. In addition, understanding phonon-phonon scattering paves the way to design strategies to tailor thermal transport. In this respect, the predictive first-principles calculations based upon solutions of the Boltzmann transport equation (BTE) for phonons \cite{Peierls-96} constitute a significant progress. These calculations allow one to calculate $\kappa_L$ without the need of adjustable parameters \cite{Ward-09,Broido-07,Li-14a}. Even though it is not clear whether this methodology is suitable for two-dimensional materials \cite{Xie-17}, its reliability in three-dimensional ones has been demonstrated for many distinct systems such as skutterudites \cite{Li-14b,Li-14c,Guo-15,Fu-16}, clathrates \cite{Tadano-15}, chalcogenides \cite{Tian-12,Carrete-14,Romero-15,Gonzalez-17} and other systems \cite{Peng-16a,Peng-16b,Mukhopadhyay-16}. An additional advantage of the first-principles calculations based upon BTE is that they shed light about heat conduction through intermediate magnitudes such as the scattering rates or the three-phonon scattering phase spaces for phonons, which are difficult to access in experiments.

\begin{figure}[h]
	\centering
	\includegraphics[width=0.9\columnwidth]{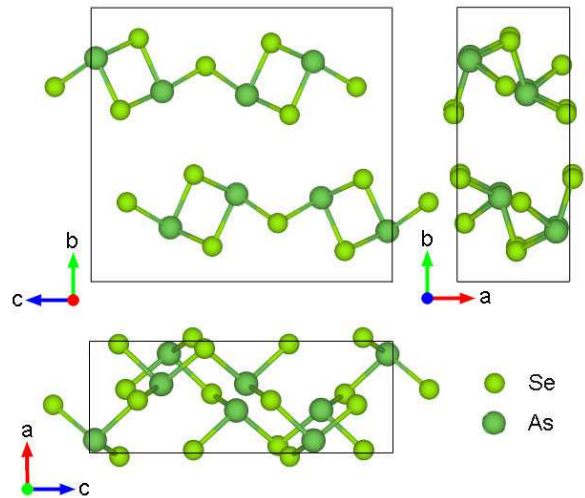}
	\caption{View of the unit cell of As$_2$Se$_3$ along the crystal axes.}
	\label{fig:as2se3}
\end{figure}

In this work, we have used a first-principles BTE approach to calculate $\kappa_L$ of As$_2$Se$_3$ single crystals. The As$_2$Se$_3$ system  was widely studied during the 80$'$s, mainly in glassy form, because of its excellent optical and electronic properties \cite{Tarnow-86a,Tarnow-86b,Struzhkin-08,Sharma-12}. As$_2$Se$_3$ is a semiconductor with relatively complex structure. It crystallizes in the monoclinic system (space group $P2_1/n$) with lattice parameters $a = 12.053$ \AA, $b = 9.890$ \AA, $c = 4.277$ \AA\ and $\beta = 90.47\degree$ \cite{Stergiou-85,Tarnow-86a}. Figure~\ref{fig:as2se3} displays the view along the crystal axes of the As$_2$Se$_3$ unit cell, which contains 20 atoms. The As atoms have two different environments, with three Se atoms surrounding them in each case. The Se atoms have three environments, with each Se bonded to two atoms. As a result, the unit cell exhibits a stacked bilayer configuration with atoms arranged in zig-zag, similar to that observed in other chalcogenides with the GeS-type structure \cite{Chattopadhyay-86}. The electronic and optical properties of As$_2$Se$_3$ have been reported elsewhere; for completeness, we will just mention that it has an indirect gap ranging between 1.3 and 1.5~eV \cite{Sharma-12,Tarnow-86a}, and its effective masses are low \cite{Tarnow-86a}. A peculiarity of As$_2$Se$_3$ is that its electronic bands are relatively flat near the gap, similar to the ``pudding mold'' bands proposed by Kuroki and Arita \cite{Kuroki-07} as suitable for optimal thermoelectrical performance. All these features make one wonder whether As$_2$Se$_3$ could be a good candidate material for potential thermoelectric applications. 

This was the main motivation of our work. A complete study of the thermoelectrical performance of this system would require to investigate the characteristics of the heat and the charge carriers transport. In this respect, the gap of As$_2$Se$_3$, as that of SnSe, is relatively large. This indicates that, in principle, its optimal carriers transport properties will appear at high temperatures, higher than its melting point (633~K), and therefore that they would have limited utility in practice. In these large-gaps semiconductors, however, the charge carrier transport is greatly affected by the existence of point defects. These may be intrinsic (i.e., vacancies), but appear mostly due to doping. In SnSe, for example, doping with small amounts of Na or Ag results in an improvement of its figure of merit \cite{Chen-14, Peng-16c}, and similar behavior could appear in As$_2$Se$_3$ as well. In any case, the preliminary step to investigate the thermoelectrical performance of this system is to evaluate its lattice thermal properties, which have not been studied before to the best of our knowledge. 

\section{Methodology}
\subsection{Theoretical framework}
Within the BTE formalism under the relaxation time approximation (RTA) and the assumption of an uniform temperature gradient \cite{Ward-09,Broido-07,Lindsay-12,Lindsay-13}, the lattice thermal conductivity at a given temperature $T$ may be calculated as a sum of contributions over all the phonon modes $\lambda$ with wave vector $\vec q$ \cite{Li-14a}:
\begin{equation}
\label{eq:kappa_L}
\kappa_L \equiv \kappa_L^{\alpha \alpha} = \frac 1{NV}\sum_{\lambda} \frac {\partial f_{\lambda}}{\partial T} (\hbar \omega_{\lambda})v_{\lambda}^{\alpha}v_{\lambda}^{\alpha}\tau_{\lambda},
\end{equation}
where $N$ is the number of uniformly spaced $\vec q$ points within the first Brillouin zone (BZ), $V$ is the volume of the unit cell, $f_{\lambda}~\equiv(\omega_{\lambda},T)$ is the Bose-Einstein distribution function, depending on the phonon frequency $\omega_{\lambda}$ and the temperature, $v_{\lambda}^{\alpha}$ is the phonon velocity along the $\alpha$ direction and $\tau_{\lambda}$ is the phonon lifetime. 

Solving BTE requires knowing the harmonic interatomic force constants (h-IFC), as input parameters to calculate the phonons frequencies and velocities, which are quantities needed to obtain the phonon lifetimes $\tau_{\lambda}$. For bulk materials, the inverse of $\tau_{\lambda}$ equals the total scattering rate, which is the sum of the isotropic $(1/\tau^{iso})$ and the anharmonic scattering rates $(1/\tau^{anh})$. In general, $1/\tau^{iso}$ is lower than $1/\tau^{anh}$ , so that only the later is usually considered in calculations of thermal conductivity (see ESI\dag, Fig.~S1, for details). On the other hand, $1/\tau^{anh}$ may be calculated as the sum of three-phonon transition probabilities $\Gamma_{\lambda \lambda^{'} \lambda^{''}}^{\pm}$, defined as \cite{Ward-09,Broido-07,Li-14a,Tian-12}:
\begin{equation}
\label{eq:Gamma}
\Gamma_{\lambda \lambda^{'} \lambda^{''}}^{\pm}=\frac {\pi\hbar}{8N} {2(f_{\lambda^{'}}-f_{\lambda^{''}}) \brace f_{\lambda^{'}}+f_{\lambda^{''}}+1} \frac {\delta(\omega_{\lambda} \pm \omega_{\lambda^{'}} - \omega_{\lambda^{''}})} {\omega_{\lambda}\omega_{\lambda^{'}}\omega_{\lambda^{''}}} |V_{\lambda\lambda^{'}\lambda^{''}}^{\pm}|^2,
\end{equation}
where the upper (lower) row within curly brackets with the ``$+$'' (``$-$'') sign corresponds to the phonon absorption (emission) processes and $\lambda$, $\lambda^{'}$ and $\lambda^{''}$ are related through their respective wave vectors as
\begin{equation}
	\vec q^{''} = \vec q \pm \vec q^{'},
	\label{eq:k_vectors}
\end{equation}
to within a reciprocal lattice vector.

In Eq.~(\ref{eq:Gamma}), $\delta$ holds for the Dirac delta distribution and $V_{\lambda\lambda^{'}\lambda^{''}}^{\pm}$ are the scattering matrix elements, which are obtained from the normalized eigenvectors for the three phonons involved in the scattering process and the corresponding anharmonic interatomic force constants (a-IFCs) \cite{Li-14a}. Both the second-order h-IFC and the third-order a-IFC can be calculated within a real-space finite-difference supercell approach within the density functional theory (DFT) \cite{Li-14a}. 

Another magnitude that sheds light about phonon scattering processes is the three-phonon scattering phase space $W^{\pm}(\omega_{\lambda})$ (the ``$+$'' and ``$-$'' signs corresponding to the absorption and emission processes, respectively), which accounts for the contribution of the harmonic phonon frequencies to the anharmonic scattering rates. For a mode $\lambda$, the three-phonon scattering phase space is defined as the sum, spanned to all modes fulfilling~(\ref{eq:k_vectors}), of the frequency-containing factors in the expression of three-phonon transition probabilities,  Eq.~(\ref{eq:Gamma}) \cite{Li-14b}:
\begin{equation}
\label{eq:w}
	W_{\lambda}^{\pm} \equiv W^{\pm} (\omega_{\lambda}) = \frac 1{2N} \sum_{\lambda^{'},\lambda^{''}} {2(f_{\lambda^{'}}-f_{\lambda^{''}}) \brace f_{\lambda^{'}}+f_{\lambda^{''}}+1} \frac {\delta(\omega_{\lambda} \pm \omega_{\lambda^{'}} - \omega_{\lambda^{''}})} {\omega_{\lambda}\omega_{\lambda^{'}}\omega_{\lambda^{''}}}
\end{equation}

\subsection{Computational methods}
DFT first-principles calculations \cite{Hohenberg-64,Kohn-65} have been performed using the projector augmented wave (PAW) method \cite{Blochl-94} as implemented in the Vienna ab initio simulation package (VASP) \cite{Kresse-96}. The Perdew-Burke-Ernzerhof functional for the generalized-gradient-approximation (GGA) was used to describe the exchange-correlation functional \cite{Perdew-96}. A van der Waals correction modeled by the DFT-D method by Grimme \textit{et al.} \cite{Grimme-10} were explicitly included, as it yields optimized results in these types of structures \cite{Gonzalez-17}. The kinetic energy cutoff of wave functions was set to 650~eV, and a Monkhorst-Pack $k$-mesh of 4$\times$6$\times$12 was used to sample BZ for integrations in the reciprocal space \cite{Monkhorst-76}. A force less than 10$^{-5}$~eV/\AA\ and a total change in energy less than 10$^{-7}$~eV were selected as convergence criteria for the structural optimization.

The first-principles lattice thermal conductivity $\kappa_L$ was calculated by solving BTE for phonons. The IFCs were calculated within a real-space supercell approach by using the Phonopy package \cite{Togo-15} for the two-order h-IFCs and the ShengBTE package \cite{Li-14a} for the third-order a-IFCs. The IFCs were calculated using a 2$\times$2$\times$2 supercell with a 2$\times$3$\times$6 $k$-mesh; the $\Gamma$ point only was used to sample BZ in this case, and a cutoff of 5.6~\AA\ for the interaction range was employed. These values yielded $\kappa_L$ values converged to within 0.1 W$\cdot \text{m}^{-1}\cdot\text{K}^{-1}$; the corresponding convergence curves are shown in ESI\dag, Fig. S2. 

\section{Results and discussion}
\subsection{Geometry optimization}
After the structural relaxation, As$_2$Se$_3$ exhibited a monoclinic unit cell with lattice parameters $a~=~ 12.20 (12.05) $ \AA, $b= 10.04 (9.89)$ \AA, $c=4.27 (4.28)$ \AA\ and $\beta =90.51 (90.47)^{\circ}$, in very good agreement with the available experimental results (into brackets) \cite{Stergiou-85,Tarnow-86a}, even better than those usually obtained within a DFT framework. In our opinion, this excellent agreement could be due to the explicit inclusion of van der Waals corrections, which have been demonstrated to be crucial in other chalcogenides \cite{Gonzalez-17}.

\begin{figure}[h]
	\centering
	\includegraphics[width=0.8\columnwidth]{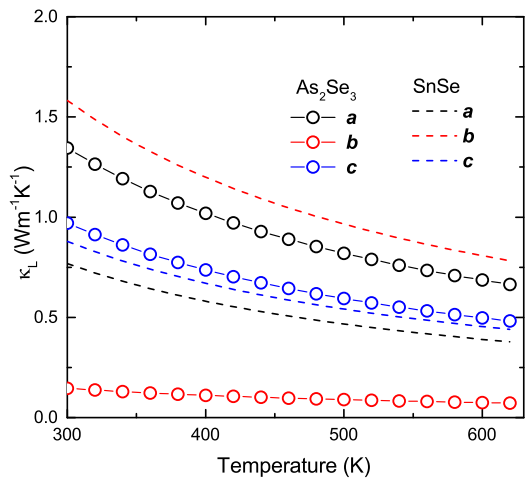}
	\caption{Lattice thermal conductivity of As$_2$Se$_3$ along the unit cell axes as functions of the temperature. Data for SnSe are included for comparison.}
	\label{fig:kappa_L}
\end{figure}

\subsection{Lattice thermal conductivity}
Figure~\ref{fig:kappa_L} shows the lattice thermal conductivity $\kappa_L$ as a function of temperature for As$_2$Se$_3$. For comparison purposes, Fig.~\ref{fig:kappa_L}  includes previous results obtained using the same methodology for SnSe as well\cite{Gonzalez-17}. This figure reveals first a strong anisotropy, with the lattice thermal conductivity along the $\vec b$ axis being roughly one order of magnitude lower than within the $ac$ plane. At room temperature, the values of $\kappa_L$ (in W$\cdot \text{m}^{-1}\cdot\text{K}^{-1}$) along the $\vec a$, $\vec b$ and $\vec c$ axes are 1.34, 0.14 and 0.97, respectively. Besides, and more importantly, Fig.~\ref{fig:kappa_L} reveals an astonishingly small value for the lattice thermal conductivity along the $\vec b$ axis. Compared to values typically reported in the literature for thermoelectric materials \cite{Zhao-14,Snyder-08}, the $\kappa_L$ value along $\vec b$ is even lower than those observed in nanostructures and all-scale hierarchical architecture of PbTe-based thermoelectrics \cite{Biswas-12}. Finally, it is remarkable that $\kappa_L$ varies with temperature as $T^{-1}$, which is the characteristic law for intrinsic three-phonon scattering. 

\begin{figure}[h]
	\centering
	\includegraphics[width=0.8\columnwidth]{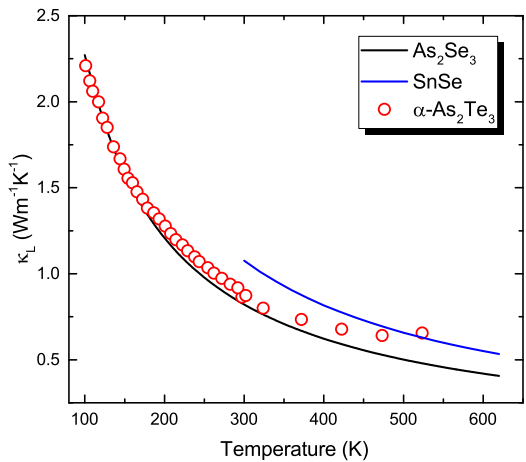}
	\caption{Comparison between our calculated lattice thermal conductivity (averaged over the three crystal axes) and experimental data for As$_2$Te$_3$ measured by Vaney \textit{et al.}\cite{Vaney-16} Data calculated for SnSe using the same methodology are also included for completeness.}
	\label{fig:as2te3}
\end{figure}

Unfortunately, there are not experimental measurements of thermal conductivity for As$_2$Se$_3$ single crystals, to our knowledge. Therefore, the validity of our results must be checked by indirect proofs. The first one comes from the comparison between our data and those available for $\alpha$-As$_2$Te$_3$ polycrystals. $\alpha$-As$_2$Te$_3$ crystallizes in the monoclinic system, with space group $C2/m$ and lattice parameters $a = 14.36$ \AA, $b = 9.90$ \AA, $c = 4.02$ \AA\ and $\beta = 95.58\degree$ \cite{Stergiou-85b}. The structure of this compound is not identical to that of As$_2$Se$_3$; in particular, the layers bonded by dispersive forces are not flat in As$_2$Te$_3$. However, they are similar enough to allow the comparison between them as an indirect sign of the validity of our calculations. Vaney \textit{et al.} found $\kappa \approx 0.6 \text{ K}\cdot\text{m}^{-1}\cdot\text{K}^{-1}$ at 550~K in $\alpha$-As$_2$Te$_3$ polycrystals \cite{Vaney-16}, while smaller values were measured for Sn-doped samples. In Figure~\ref{fig:as2te3}, the experimental data for As$_2$Te$_3$ by Vaney \textit{et al.} are plotted together with our values, calculated as an average over the three crystal axes, showing a very good agreement in the 100~K - 300~K temperature range. Sharma and Srivastava \cite{Sharma-11}, on the other hand, calculated the thermal conductivity of $\alpha$-As$_2$Se$_3$ as 0.7 $\text{K}\cdot\text{m}^{-1}\cdot\text{K}^{-1}$. We notice that, in each case, the authors report the total thermal conductivity, which can be considered as an upper limit to the lattice thermal conductivity considered herein. In any case, these results point out that monoclinic but quasi-orthorhombic systems like As$_2$Te$_3$ may exhibit low thermal conductivities. 

More importantly, the reliability of our methodology may be proved by previous papers which have used it successfully to study the thermoelectric behavior of semiconductors \cite{Fu-16,Li-15,Li-14a,Li-14b,Carrete-14,Gonzalez-17}. Using the same methodology described above, Li \textit{et al.} \cite{Li-14a} found a good agreement (to within $\pm$ 7\%) between the calculated lattice thermal conductivity of InAs (with zinc-blende structure) and experimental data by LeGuillou and Albany \cite{Leguillou-72}. Li and Mingo \cite{Li-14b} used again the same methodology to describe the lattice thermal conductivity of unfilled CoSb$_3$ and IrSb$_3$ skutterudites, which are systems with relatively complex structures. Here again, there is an excellent agreement between their calculated data and experimental data in a wide temperature range (100 - 700~K); the small differences were attributed to scattering by impurities, not included in the calculation. Finally, the above methodology has been used to explain the thermoelectrical behavior of SnSe, a system with promising properties due to its high figure of merit \cite{Zhao-14b}. Carrete \textit{et al.} \cite{Carrete-14} and Gonz\'alez-Romero \textit{et al.} \cite{Gonzalez-17} have used it to describe the phonon contribution to the thermal conductivity in SnSe, in both cases with good agreement with available experimental data. 

\subsection{Analysis of the phonon spectrum}
\begin{figure}[h]
	\centering
	\includegraphics[width=\columnwidth]{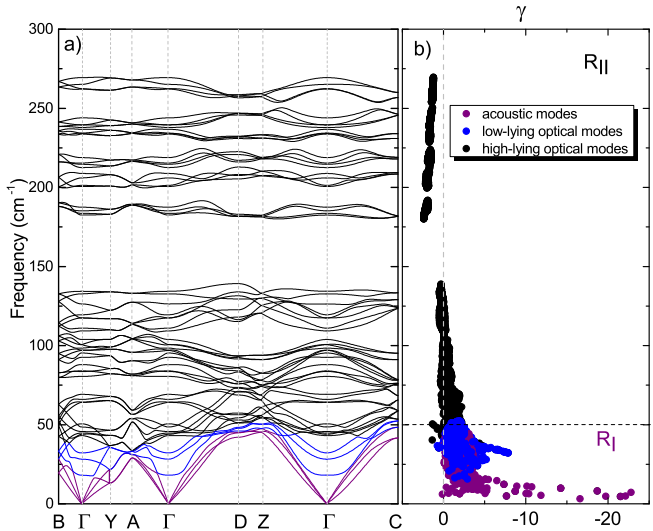}
	\caption{a) Phonon spectrum for As$_2$Se$_3$ along high-symmetry directions of BZ. b) Variation of the Gr\"{u}neisen parameter with the phonon frequency. }
	\label{fig:phonon}
\end{figure}

As we mentioned in the introduction, the deep comprehension of the thermal conductivity phenomena in semiconductors requires a correct description of their vibrational behavior. Fig.~\ref{fig:phonon}a plots the phonon spectrum of As$_2$Se$_3$ along high-symmetry directions of the BZ. This spectrum is moderately complex, with a two-part configuration containing a wide gap between 140 - 180 cm$^{-1}$, approximately, and four narrower gaps within the 180 - 275 cm$^{-1}$ range. In addition to this, the phonon spectrum shows two remarkable characteristics. The first one is the severe mixture of low-lying optical modes (LLO, blue lines in Fig.~\ref{fig:phonon}a) and acoustic modes (purple lines). For this particular case, the LLO branches exhibit frequencies of 18 (21.5), 28 (27) and 32~(32.5)~cm$^{-1}$ at $\Gamma$, in good agreement with Raman spectroscopy results reported by Zallen and Slade (into brackets) \cite{Zallen-74}. The cascade-like fall of the optical modes appears in a number of solids with very low thermal conductivities \cite{Romero-15,Lee-14,Luo-16,Delaire-11}. 

\begin{figure}[!h]
	\centering
	\includegraphics[width=\columnwidth]{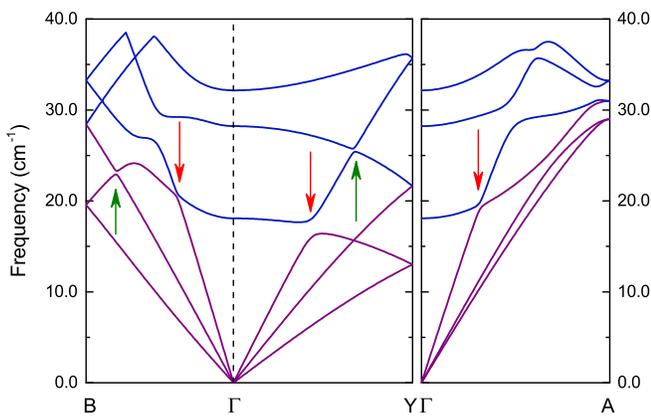}
	\caption{Detail of the low-energy (< 4.9 meV) of the phonon spectrum of As$_2$Se$_3$ along the $\Gamma-Y$, $\Gamma-B$ and $\Gamma-A$ directions. Symbols highlight the points of avoided crossing. The arrows highlight avoided-crossing points between the acoustic and optical modes (red arrows) or within the acoustic or optical branches manifolds (green arrows). The color code in this figure is the same as in Fig.~\ref{fig:phonon}.}
	\label{fig:avoided}
\end{figure}

The second relevant feature is that the phonon spectrum shows avoided crossing behavior of the acoustic and the LLO modes around along several crystallographic directions. This is evidenced in Fig.~\ref{fig:avoided}, which shows the low-energy (< 40 cm$^{-1}$ = 4.9 meV) region of the spectrum along the $\Gamma-Y (\xi,0,0)$, $\Gamma-B (0,0,\xi)$ and $\Gamma-A (\xi, 0, \xi)$ directions; the avoided-crossing points between the acoustic and LLO branches are highlighted by red arrows. From this figure, one observes that the gap at the avoided-crossing point varies with the direction within BZ; in particular, it is more important along $\Gamma - Y$ than along the other directions. The avoided crossing of two branches arises as a result of the coupling between them, whose strength is related to the gap at the avoided-crossing point. This coupling causes the hybridization of the modes in the two branches (acoustic and optical in our case). In particular, the hybridization is maximum at the $\vec k$ point at which the gap opens. At this point, both the acoustic and optical modes contribute equally to the eigenvector of the dynamical matrix giving the phonon polarization, so that these modes are no longer distinguishable \cite{Li-16}. Incidentally, we found avoided-crossing of acoustic modes with different polarizations along $\Gamma-B$ and of optical ones along $\Gamma-Y$, highlighted by green arrows in Fig.~\ref{fig:avoided}. These avoided crossing are indicative of modes with different symmetries. In this case, the analysis of the polarizations of the modes is difficult because of the monoclinic symmetry, so that we will not discuss it further. 

This avoided-crossing phenomenon (or, alternatively, the hybridization of the corresponding modes) and the mixture of acoustic and LLO modes in the low-energy region of the spectrum affect the thermal conductivity in two ways, which are correlated to each other \cite{Norouzzadeh-17}. First, they reduce the Debye temperature $\theta_D$ due to the shrinkage of the phonon bandwidth. Indeed, the presence of LLO phonons results in a softening of the acoustic branches, which yields low sound velocities and, therefore, low $\theta_D$. From the phonon spectrum in Fig.~\ref{fig:phonon}a one gets $\theta_D=128$~K, $\theta_D=116$~K and $\theta_D=181$~K along the $\vec a$, $\vec b$ and $\vec c$ crystal axes, respectively, which correlates well with the strong anisotropy of the system. According to the Slack theory \cite{Morelli-06}, low Debye temperatures are indicative of low thermal conductivities. The reason is that low $\theta_D$ values are related to low group velocities and frequencies for acoustic phonons \cite{Peng-16b}. Besides, the avoided-crossing and mixture of modes causes the phonon scattering rates to increase by reducing the phonon lifetimes. This effect is important in our case, and will be discussed in next section.

\subsection{Phonon scattering rates}
The phonon scattering rate depends basically upon two factors \cite{Peng-16b}: the strength of each scattering channel, which depends on the anharmonicity and is described by the Gr\"{u}neisen parameter $\gamma$, and the number of channels available for the phonons to be scattered; the later are characterized by the phase spaces for three-phonon processes ($W^{\pm}_{\lambda}$). Fig.~\ref{fig:phonon}b shows the Gr\"{u}neisen parameter of As$_2$Se$_3$ as a function of frequency. For clarity, we have divided the plot into two regions, namely $R_I$, corresponding to acoustic and LLO phonons, and $R_{II}$, corresponding to optical modes with frequency above 50~cm$^{-1}$. The avoided-crossing and phonon softening result in an increase of the anharmonicity of the system, and $\gamma$ reaches high absolute values. Hence, it is expected that the system exhibits low thermal conductivity, as has been observed in other systems \cite{Tian-12,Delaire-11}. In addition, according to Fu \textit{et al.} \cite{Fu-16}, negative $\gamma$ values for the LLO branches imply that these modes may be softened further by compression and, therefore, may exhibit a more severe overlap with the acoustic branches, as well as wider $W^{\pm}$ values. In other words, it is expectable that $\kappa_L$ decreases even more at high pressures. This hypothesis is well beyond the scope of the present work, but seems an interesting point to be explored further. 

\begin{figure}[h]
	\centering
	\includegraphics[width=0.8\columnwidth]{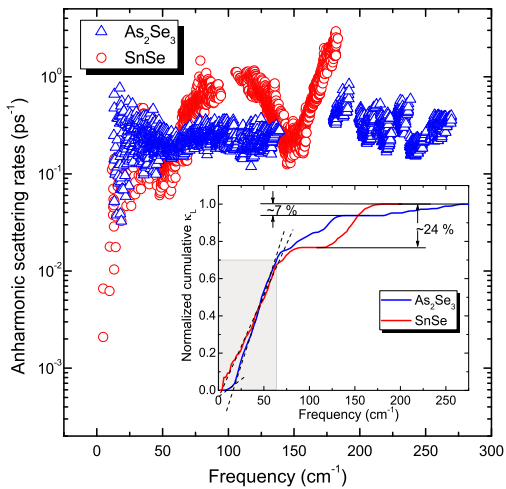}
	\caption{Anharmonic scattering rates vs. frequency for As$_2$Se$_3$ and SnSe, calculated at 300~K. The inset shows the cumulative lattice conductivity as a function of the frequency for both systems.}
	\label{fig:rates}
\end{figure}

In Eq.~(\ref{eq:kappa_L}) we show that the phonon lifetimes are directly related with $\kappa_L$. At temperatures above $\theta_D$, they are ruled by three-phonon anharmonic scattering processes \cite{Fu-16}. Fig.~\ref{fig:rates} plots the anharmonic scattering rates $1/\tau^{anh}$, where $\tau^{anh}$ is the mean anharmonic lifetime for phonons, calculated at 300~K. For comparison purposes, results for SnSe are included as well \cite{Gonzalez-17}. The inset in Fig.~\ref{fig:rates} shows the normalized cumulative lattice thermal conductivity vs. the phonon frequency for As$_2$Se$_3$ and SnSe. As we have already mentioned, the latter has been successfully described using the present methodology. Thence, in absence of experimental data for As$_2$Se$_3$, we will check the plausibility of our results by comparing them with those for SnSe. 

In both systems, the anharmonic scattering rates for acoustic and optical phonons with frequencies below approximately 50~cm$^{-1}$ are between two (As$_2$Se$_3$ case) and three (SnSe case) orders of magnitude lower than those with higher frequencies, for each material. For As$_2$Se$_3$, the scattering rates for ultralow-frequency phonons are higher than for SnSe; the contribution of the latter to $\kappa_L$ is obviously different, as evidenced by the different slopes in the inset of Fig.~\ref{fig:rates}. In addition, in both systems, roughly 70\% of the heat is driven by phonons with frequency below 65~cm$^{-1}$ (shaded region in the inset of Fig.~\ref{fig:rates}). In this same frequency range one observes a split between the scattering rates for both systems; thus, at around 65~cm$^{-1}$ the scattering rates are approximately equal for As$_2$Se$_3$ as for SnSe, but at higher frequencies $1/\tau^{anh}$ is smaller for the former. It is worthwhile to note that the contribution to the heat current by the optical branches with frequencies above the gap is different in both cases. From the inset in Fig.~\ref{fig:rates} one observes that the contribution of these modes is around 24\% for SnSe, whereas it is barely of the 7\% for As$_2$Se$_3$.

\begin{figure}[h]
	\centering
	\includegraphics[width=0.8\columnwidth]{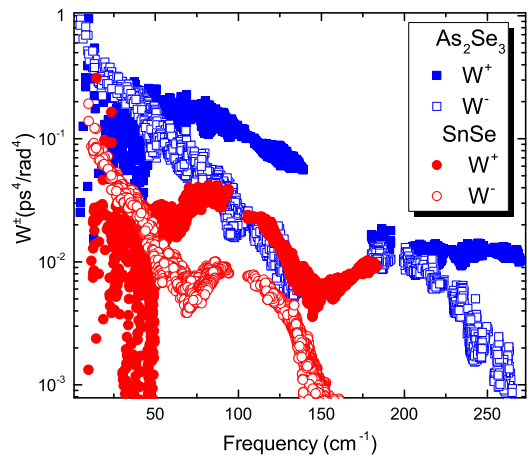}
	\caption{Absorption and emission three-phonon scattering phase spaces for As$_2$Se$_3$ and SnSe as functions of the frequency. }
	\label{fig:W}
\end{figure}

According to Eq.~(\ref{eq:Gamma}), the anharmonic scattering rates are proportional to the squares of the a-IFCs. Therefore, the anharmonic interatomic force constants for both systems are approximately comparable, at least for frequencies above about 30~cm$^{-1}$. The question arises as to understand the low conductivity of As$_2$Se$_3$ along the $\vec b$ axis, that is, perpendicularly to the structural bilayers. To clarify this issue, Fig.~\ref{fig:W} plots the three-phonon scattering phase spaces, $W^{\pm}_{\lambda}$. Essentially, these quantities represent the number of scattering channels for three-phonon processes available to all the modes, and vary inversely with $\kappa_L$ \cite{Nielsen-13}. Thus, the $W^{\pm}_{\lambda}$ values provide insights about the effect of the phonon frequencies on the anharmonic scattering rate via their contribution to the three-phonon scattering matrix elements \cite{Fu-16}.

Fig.~\ref{fig:W} indicates that the scattering phase spaces are significantly higher in As$_2$Se$_3$, especially at low vibrational frequencies, whereas both As$_2$Se$_3$ and SnSe exhibit comparable values of $W^{\pm}_{\lambda}$ for optical modes with frequencies above their respective gaps. At low frequencies, both absorption and emission are equally probable in As$_2$Se$_3$, as it was expectable. However, there are marked differences between 50 and 150~cm$^{-1}$, as well as at frequencies above about 220~cm$^{-1}$, where absorption predominates. The widening of $W^{\pm}_{\lambda}$ for As$_2$Se$_3$ results from the shape of the phonon spectrum, where the LLO branches overlap the acoustic ones (see Fig.~\ref{fig:phonon}a). As a consequence, the anharmonicity of the system as well as the phase spaces increase, and this facilitates phonon scattering.

\begin{figure}
	\centering
	\includegraphics[width=0.9\columnwidth]{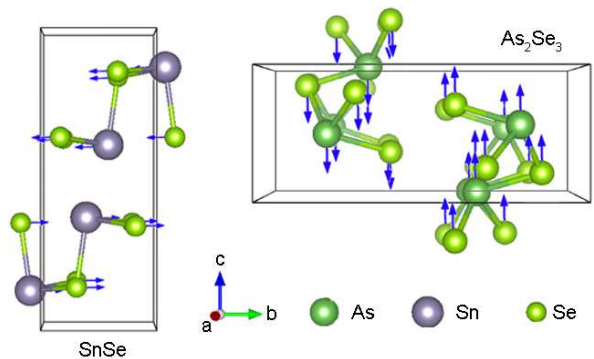}
	\caption{Schematic representation of the low-lying optical modes causing the rigid vibration of the atomic layers along the $\vec b$ (SnSe, left figure) and $\vec c$ (Al$_2$Se$_3$, upper right figure) axes.}
	\label{fig:vibrations}
\end{figure}

Two mechanisms are at play to lower the lattice thermal conductivity of layered chalcogenide materials, such as As$_2$Se$_3$ and SnSe: 1) the avoided crossing between the acoustic and LLO modes due to their hybridization, which softens the acoustic branches, resulting in low sound velocities, and increases the scattering rate of the acoustic phonons (previously discussed); and 2) the mechanism proposed by Nielsen \textit{et al.}\cite{Nielsen-13}, by which the repulsion between the lone-pair electrons of cations of the groups IV and V and the $p$ orbitals of the chalcogenide anions leads to an increase in the anharmonicity of the covalent bonds. The anisotropy in the lattice thermal conductivity is related to the anisotropy in the chemical bonding. In As$_2$Se$_3$, the intralayer bonding is covalent, while the interlayer bonds have a strong van der Waals character. In the $Pnma$ phase of SnSe, the chemical bonding is constituted by resonant covalent bonds, which are stronger within the layers, while the interlayer bonds are weaker, also exhibiting van der Waals character \cite{Li-15b}. The first mechanism is directly connected with the anisotropy in the chemical bonding of these materials, while the second one is primarily responsible for the low lattice thermal conductivity within the layers. Unfortunately, our calculations are unable to separate the individual contributions of each mechanism. The low frequency of the LLO modes is caused by two factors, namely the weak interlayer bonds and the essentially rigid motion of the layers in these modes (large mass), as depicted in Fig.~\ref{fig:vibrations}. The avoided crossings in As$_2$Se$_3$ lie about the frequency of 20 cm$^{-1}$, while in SnSe the avoided crossings appear at higher frequencies, above 30 cm$^{-1}$ \cite{Carrete-14,Gonzalez-17}. This significant discrepancy if possibly due to the difference in the strength of the interlayer bonds, which are weaker in As$_2$Se$_3$ than in SnSe. This suggests that the effect of the avoided crossings on the lattice thermal conductivity may be stronger in As$_2$Se$_3$ than in SnSe, thus explaining the ultralow lattice thermal conductivity along the $\vec b$ direction found in this work.

\section{Conclusions}
Using first-principles methods (namely DFT and the BTE formalisms), we have shown that As$_2$Se$_3$ exhibits a low lattice thermal conductivity, of the order of 0.14~W$\cdot\text{m}^{-1}\cdot\text{K}^{-1}$ at room temperature, along the $\vec b$ axis. This anisotropic low thermal conductivity is related, on one hand, to the strong anharmonicity of the system and, on the other hand, to its high scattering phase space values. Both features may be understood, in turn, on the basis of two mechanisms, namely the interaction between the acoustic and LLO modes, which reduces the sound velocities and increase the scattering rates of the acoustic phonons, and the repulsion between lone-pair electrons of As and valence $p$ orbitals of Se, which enhances the anharmonicity of the chemical bonds. The ultralow lattice thermal conductivity along the $\vec b$ axis of As$_2$Se$_3$ stems from its electronic structure and the resulting anisotropic chemical bonding, which lead to exceptionally low-frequency avoided crossings between acoustic and LLO modes. Our results suggest that As$_2$Se$_3$ could be a promising candidate for thermoelectric applications which deserves an experimental exploration. 

\section{Acknowledgements}
We gratefully acknowledge support from the Brazilian agencies CNPq, CAPES, and FAPESP under Grants \#2013/14065-7, \#2013/08293-7 and \#2015/26434-2. Financial support by the Junta of Extremadura in Spain through Grants GR15105 and IB16013 is acknowledged as well. The calculations were performed at CCJDR-IFGW-UNICAMP and at CENAPAD-SP in Brazil.

\bibliography{as2se3}
\bibliographystyle{rsc}

\balance
\end{document}